\documentclass[superscriptaddress,twocolumn,prb,showpacs]{revtex4}
\usepackage[dvips]{graphicx}
\usepackage{amsmath}

\renewcommand{\[}{\begin{equation}}
\renewcommand{\]}{\end{equation}}

\newcommand{\equ}[1]{Eq.~(\ref{#1})}
\newcommand{\eqs}[2]{Eqs.~(\ref{#1}) and (\ref{#2})}
\newcommand{\ei}[1]{{\rm e}^{i #1}}
\newcommand{\emi}[1]{{\rm e}^{-i #1}}

\newcommand{\intk}{\int \! d{\bf k} \;}

\newcommand{\da}{\partial_\alpha}
\newcommand{\db}{\partial_\beta}
\newcommand{\kk}{\mbox{\boldmath$\kappa$}}

\def\beq{\begin{equation}}
\def\eeq{\end{equation}}
\def\bea{\begin{eqnarray}}
\def\eea{\end{eqnarray}}

\def\ket#1{\vert#1\rangle}
\def\bra#1{\langle#1\vert}
\def\me#1#2#3{\langle#1\vert#2\vert#3\rangle}
\def\ev#1{\langle#1\rangle}
\def\r{{\bf r}}

\def\v{{\bf v}}
\def\k{{\bf k}}

\begin{document}
%%%%%%%%%%%%%%%%%%%%%%%%%%%%%%%%%%%%%%%%%%%%%%%%%%%%%%%%%%%%%%%%%%%%%%%%%%%

\title{Orbital magnetization and Chern number in a supercell framework: \\
Single {\bf k}-point formula}

\author{Davide Ceresoli}
\affiliation{Scuola Internazionale Superiore di Studi Avanzati (SISSA)
and CNR-INFM Democritos National Simulation Center, via Beirut 2-4, 34014 Trieste, Italy}
\author{Raffaele Resta}
\affiliation{Dipartimento di Fisica Teorica, Universit\`a di Trieste, and
CNR-INFM Democritos National Simulation Center, strada Costiera 11, 34014 Trieste, Italy}
\date{\today}
%\date{\sf DRAFT: run through \LaTeX\ on \today\ at \hour}

%%%%%%%%%%%%%%%%%%%%%%%%%%%%%%%%%%%%%%%%%%%%%%%%%%%%%%%%%%%%%%%%%%%%%%%
\begin{abstract}
%%%%%%%%%%%%%%%%%%%%%%%%%%%%%%%%%%%%%%%%%%%%%%%%%%%%%%%%%%%%%%%%%%%%%%%

The key formula for computing the orbital magnetization of a crystalline system
has been recently found [D. Ceresoli, T. Thonhauser, D. Vanderbilt, R. Resta,
Phys. Rev. B {\bf 74}, 024408 (2006)]: it is given in terms of a Brillouin-zone
integral, which is discretized on a reciprocal-space mesh for numerical
implementation. We find here the single ${\bf k}$-point limit, useful for large
enough supercells, and particularly in the framework of Car-Parrinello
simulations for noncrystalline systems. We validate our formula on the test case
of a crystalline system, where the supercell is chosen as a large multiple of
the elementary cell. We also show that---somewhat counterintuitively---even the
Chern number (in 2d) can be evaluated using a single Hamiltonian
diagonalization.

\end{abstract}
%%%%%%%%%%%%%%%%%%%%%%%%%%%%%%%%%%%%%%%%%%%%%%%%%%%%%%%%%%%%%%%%%%%%%%%%
% 75.10.-b 	General theory and models of magnetic ordering
% 75.10.Lp 	Band and itinerant models
% 73.20.At 	Surface states, band structure, electron density of states
% 73.43.-f 	Quantum Hall effects
%%%%%%%%%%%%%%%%%%%%%%%%%%%%%%%%%%%%%%%%%%%%%%%%%%%%%%%%%%%%%%%%%%%%%%%%
\maketitle

The position operator ${\bf r}$ is ill-defined within periodic boundary
conditions. Owing to this, both the macroscopic (electrical) polarization and
the macroscopic orbital magnetization are nontrivial quantities in
condensed-matter theory. The former has been successfully tamed since the
early 1990s, when the modern theory of polarization, based on a Berry phase,
was developed.\cite{KSV,rap_a12} The latter, instead, remained an unsolved
problem until 2005. Since then, several important papers have
appeared\cite{rap126,Xiao05,rap128,rap130,Xiao07} and continue to
appear.\cite{Shi07} Before 2005 only linear-response properties related to
orbital magnetization were successfully addressed,\cite{linear,Mauri} while
we stress that the present work, as well as 
Refs.~\onlinecite{rap126,Xiao05,rap128,rap130,Xiao07,Shi07}, addresses 
``magnetization itself''.

A general formula, valid for crystalline systems within a given single-particle
Hamiltonian, was provided in Ref.~\onlinecite{rap130}, hereafter referred to as
I. This is the magnetic analogue of the (by now famous) King-Smith and
Vanderbilt formula for electrical polarization.\cite{KSV}  Both formulas are
discretized on a regular mesh of ${\bf k}$ points for numerical implementation. 
However, most simulations for noncrystalline systems, particularly those of the
Car-Parrinello type,\cite{CP} are routinely performed by diagonalizing the
Hamiltonian at a single ${\bf k}$ point (the $\Gamma$ point) in a large
supercell. The reduction from many points to a single point is far from being
trivial; nonetheless a successful single-point formula for electrical
polarization emerged since 1996, and is universally used since
then.\cite{trois,rap100}  We provide here the magnetic analogue of such formula.
As a byproduct, one can even evaluate the Chern number~\cite{Thouless} using a
single ${\bf k}$ point. This looks like an oxymoron, given that the Chern
number is by definition a loop integral in reciprocal space: but our formula can
be regarded as the limiting case where the loop shrinks to a single point.

The general formula of I applies to normal
periodic insulators (where the Chern invariant vanishes), Chern insulators
(where the Chern invariant is nonzero), and metals.  Aiming at
first-principle implementations within any flavor of DFT, the single-particle
Hamiltonian is the Kohn-Sham one.\cite{DFT} As for the analogous
case of electrical polarization, there is no guarantee that the Kohn-Sham
magnetization coincides with the actual one. Nonetheless, previous studies
within linear-response methods indicate that even for orbital magnetization
the error is small.\cite{Mauri}

As in I, we assume a vanishing macroscopic magnetic field ${\bf B}$, hence a
lattice-periodical Hamiltonian. We let $\epsilon_{n\k}$ and $\ket{\psi_{n\k}}$
be the Bloch eigenvalues and eigenvectors of $H$, respectively, and
$u_{n\k}(\r)=\emi{\k\cdot\r}\psi_{n\k}(\r)$ be the corresponding eigenfunctions
of the effective Hamiltonian \beq H_\k = \emi{\k\cdot\r} H \ei{\k\cdot\r} \, ;
\label{Hk} \eeq  we normalize them to one over the crystal cell of volume $V$.
As in I, the  notation is intended to be flexible as regards the spin character
of the electrons. If we deal with spinless electrons, then $n$ is a simple index
labeling the occupied Bloch states; factors of two may trivially be inserted if
one has in mind degenerate, independent spin channels. For the sake of
simplicity, we rule out the metallic case here.  For both normal insulators and
Chern insulators the macroscopic orbital magnetization ${\bf M}$ is, according
to I: \bea M_{\gamma} &=& -\frac{\varepsilon_{\gamma\alpha\beta}}{2c(2\pi)^3 }
\label{M} \\ \nonumber  &\times& {\rm Im}\,   \sum_{n} \intk \bra{\da u_{n\k}}
(\, H_\k + \epsilon_{n\k} -2 \mu \,)\,\ket{\db u_{n\k}}\;,  \eea where Greek
subscripts are Cartesian indices, $\varepsilon_{\gamma\alpha\beta}$ is the
antisymmetric tensor, $\da = \partial/\partial k_\alpha$, $\mu$ is the chemical
potential (Fermi energy), the integration is over the reciprocal cell, and the
sum over Cartesian indices is implied. For insulators the number of states with
energy $\epsilon_{n\k} \leq \mu$ is independent of $\k$; we implicitly
understand the sum in \equ{M} as on these states only.  

As usual, a noncrystalline system can be  dealt with in a supercell framework,
by addressing an artificial crystal with a large unit cell, actually larger than
the relevant correlation length. One key feature of \equ{M} is its
gauge-invariance in the generalized sense: by this we mean that \equ{M} is
invariant by unitary mixing of the occupied states among themselves at a any
given $\k$. Thanks to such key feature \equ{M} is invariant by cell doubling. In
fact, starting with a cell (or supercell) of given size, we may regard the same
physical system as having double periodicity (in all directions), in which case
the integration domain in \equ{M} gets ``folded'' and shrinks by a factor $1/8$,
while the number of occupied eigenstates gets multiplied by a factor of $8$. It
is easy to realize that these are in fact {\it the same} eigenstates as in the
unfolded case, apart possibly for a unitary transformation, irrelevant here.  As
for actual computations, the discretized form of \equ{M} adopted in I turns to
be invariant by cell doubling to within $10^{-5}$, provided the ${\bf k}$-point
mesh is chosen consistently.

For a large supercell of volume $V$  the integration domain in
\equ{M} becomes small. Therefore the integral can be accurately
approximated by the value of the integrand at $\k =0$, times the
reciprocal-cell volume: \beq M_{\gamma} \simeq
-\frac{\varepsilon_{\gamma\alpha\beta}}{2c V }\, \,{\rm Im}\, \sum_n  \bra{\da
u_{n0}} (\, H_0 + \epsilon_{n0} -2 \mu \,)\,\ket{\db u_{n0}}\;.  \label{mean}
\eeq Notice that \equ{M} can be safely approximated with \equ{mean} because
the {\it integrand} is a gauge-invariant quantity; the apparently analogous
case of polarization is more difficult, since the integrand therein is
gauge-dependent, and the single-point formula requires a less straightforward
treatment.\cite{trois,rap100} As for the derivatives in \equ{mean}, they can
be evaluted in two ways: either ``analytically'' (by means of
perturbation theory), or ``numerically'' (by means of finite differences).

The analytical-derivative approach relies on the perturbation formula \beq
\ket{\da u_{n0}} = \sum_{m \neq n} \ket{u_{m0}}
\frac{\me{u_{m0}}{v_\alpha}{u_{n0}}}{\epsilon_{m0} - \epsilon_{n0}} \;
,\label{pert} \eeq where $v_\alpha$ is the $\alpha$-component of the velocity
operator \beq \v = i[H,\r] = \left. \nabla_\k H_\k  \right|_{\k=0} \; .
\label{v} \eeq \equ{pert} is convenient for tight-binding implementations,
where the sum is over a small number of terms. We also notice that the 
matrix representation of $\r$, for use in \eqs{Hk}{v}, is usually taken to be
diagonal on the tight-binding basis.

The numerical-derivative approach looks more convenient for first-principle
implementations, since it requires neither the (slowly convergent) sum over
states, nor the matrix elements of the velocity operator. In order to give
the most general formulation for non-rectangular cells it is expedient to
switch to a coordinate-independent form. In order to shorten the equations,
in all of the following developments of \equ{mean} a sum over the occupied
states is implied.

If ${\bf b}_j$ are the shortest reciprocal vectors
of the supercell,  \equ{mean} can be identically recast as \beq {\bf M} \simeq
-\frac{\varepsilon_{ijl} {\bf b}_i |{\bf b}_j| |{\bf b}_l| }{2c (2\pi)^3} \,{\rm
Im}\, \bra{\partial_j u_{n0}} (\, H_0 + \epsilon_{n0} -2 \mu
\,)\,\ket{\partial_l u_{n0}} .  \label{mean2} \eeq  where a sum over $ijl$ is
implied, and $\partial_j$ indicates the partial derivative in the direction of
${\bf b}_j$: \beq \ket{\partial_j u_{n0}} = \lim_{\lambda \rightarrow 0} \;
\frac{1}{\lambda |{\bf b}_j|} ( \, \ket{u_{n\,\lambda{\bf b}_j }} - \ket{u_{n0}}
\, ) . \label{der} \eeq  Notice that \equ{der} implicitly assumes that
$\ket{u_{n\k}}$ is a differentiable function: but this is generally  {\it not}
the case when the eigenstates are obtained from numerical diagonalization. The
discretization must then be done in a specific gauge: as in I, we fix the
problem by adopting the ``covariant derivative'' approach, introduced in
Refs.~\onlinecite{Sai02} and~\onlinecite{Souza04}. One defines the overlap matrix $ S_{nn'}(\k) =
\ev{u_{n0} | u_{n'\k}} $, and the ``dual'' states \beq \ket{\tilde{u}_{n\k}} =
\sum_{n'} S^{-1} _{n'n}(\k) \, \ket{u_{n'\k}} , \label{dual} \eeq which enjoy the property
$\ev{u_{n0} | \tilde{u}_{n'\k}} = \delta_{nn'}$. Using this, approximating
\equ{der} with its $\lambda =1$ value, and inserting in \equ{mean2} we finally
get \beq {\bf M} \simeq -\frac{\varepsilon_{ijl} {\bf b}_i }{2c (2\pi)^3} \,{\rm
Im}\, \bra{\tilde{u}_{n{\bf b}_j}} (\, H_0 + \epsilon_{n0} -2 \mu
\,)\,\ket{\tilde{u}_{n{\bf b}_l}} . \label{mean3} \eeq Next, we wish to evaluate
$\ket{\tilde{u}_{n{\bf b}_j}}$ without actually diagonalizing the Hamiltonian at
$\k \neq 0$. To this aim, we notice that  the state $\emi{{\bf b}_j \cdot \r}
\ket{u_{n0}}$ obeys periodic boundary conditions and is an eigenstate of
$H_{n{\bf b}_j}$ corresponding, possibly, to a different occupied eigenvalue.
The transformation in \equ{dual} restores the correct ordering anyhow; therefore
we can simply identify $\ket{u_{n{\bf b}_j}} = \emi{{\bf b}_j \cdot \r}
\ket{u_{n0}}$, transform to the dual states by means of \equ{dual}, and insert
into \equ{mean3} which eventually becomes the single-point, ${\bf k} = 0$
formula, aimed at.

For a two-dimensional (2d) system the magnetization is a pseudoscalar, and the
analogue of \equ{mean3} reads \beq M = -\frac{|{\bf b}_1 \times {\bf
b}_2|}{c(2\pi)^2|{\bf b}_1| \, |{\bf b}_2|} \,{\rm Im}\,
\bra{\tilde{u}_{n{\bf b}_1}} (\, H_0 + \epsilon_{n0} -2 \mu
\,)\,\ket{\tilde{u}_{n{\bf b}_2}} . \label{mean4} \eeq Similarly,
the single-point formula for the Chern number reads
\beq C = -\frac{|{\bf b}_1 \times {\bf
b}_2|}{2\pi |{\bf b}_1| \, |{\bf b}_2|} \,{\rm Im}\,
\ev{\tilde{u}_{n{\bf b}_1} | \tilde{u}_{n{\bf b}_2}} . \label{chern} \eeq

\begin{figure}\begin{center}
  \includegraphics[width=0.8\columnwidth]{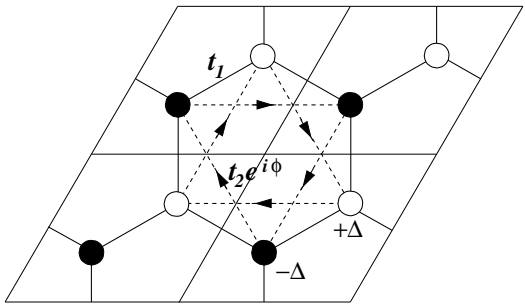}
  \caption{Four unit cells of the Haldane model.
  Filled (open) circles denote sites with $E_0=-\Delta$ ($+\Delta$).
  Solid lines connecting nearest neighbors indicate a real hopping
  amplitude $t_1$; dashed arrows pointing to a second-neighbor site
  indicates a complex hopping amplitude $t_2e^{i\phi}$.
  Arrows indicate sign of the phase $\phi$ for second-neighbor hopping.}
  \label{fig:haldane}
\end{center}\end{figure}

As in previous works,~\cite{rap128,rap130} we find expedient to validate the
present findings on the Haldane model Hamiltonian:\cite{Haldane88} it is
comprised of a 2d  honeycomb lattice with two tight-binding sites per primitive
cell with site energies $\pm \Delta$, real first-neighbor hoppings $t_1$, and
complex second-neighbor hoppings $t_2e^{\pm i\varphi}$, as shown in
Fig.~\ref{fig:haldane}. Within this two-band model, one deals with insulators by
taking the lowest band as occupied.  Following the original
notations\cite{Haldane88}  we choose the parameters $\Delta = 1$, $t_1 = 1$ and
$|t_2| = 1/3$. As a function of the flux parameter $\phi$, this system undergoes
a transition from zero Chern number to $|C|=1$ when $|\sin\phi| > 1/\sqrt{3}$.
Here we address periodic supercells made of $L \times L$ primitive cells, up to
$L=32$ (2048 sites), taking the lowest $L^2$ orbitals as occupied.

\begin{figure}
\includegraphics[angle=-90,width=0.8\columnwidth]{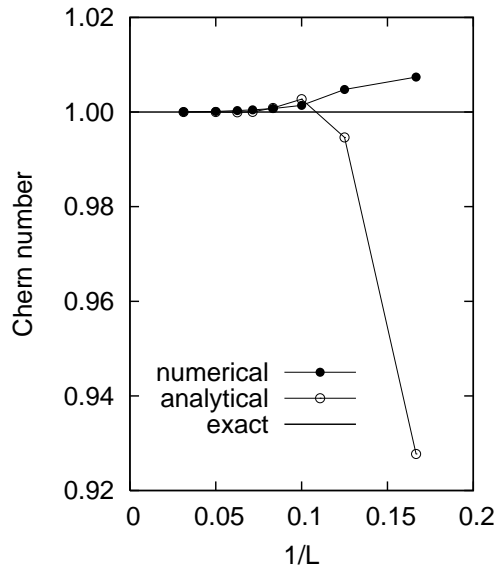}
\caption{Convergence of the Chern number as a function of the supercell size,
evaluated using the single-point formulas (see text), for the Haldane model
Hamiltonian at $\phi = 0.4\pi$. The largest $L$ corresponds to 2048 sites.}
\label{fig:f2} \end{figure}

Before actually addressing magnetization, we start benchmarking the accuracy of
our single-point formula for the Chern number, whose value is known exactly as a
function of the parameters of the model. The convergence of the Chern
number---computed from \equ{chern} and its analytical-derivative analogue---as a
function of the supercell size, is shown in Fig.~\ref{fig:f2}, for $\phi = 0.4 \pi$, where
the exact value is 1. Both approaches (analytical and numerical derivative)
converge very fast. For instance the numerical-derivative approach yields an
error of $7 \times 10^{-3}$ for $L=6$, and smaller than  $10^{-5}$ for $L=32$.
We are showing here the results for a $\phi$ value well inside the $C=1$
domain. We also find that the convergence worsens near the transition point
$|\sin\phi| = 1/\sqrt{3}$. 

Numerical evaluation of Chern numbers is a staple tool in the theory of the
quantum Hall effect, where supercells are routinely used to account for disorder
and/or electron-electron interaction. However, even in a supercell framework, a
discrete reciprocal mesh (or equivalently a mesh of phase boundary conditions)
has been invariably used in the algorithms implemented so
far.\cite{Yang96,Yang99,Sheng03,Wan05} Here we have shown that, provided the
supercell is large enough, no mesh is needed: the Chern number can be evaluated
from a single Hamiltonian diagonalization (with a single choice of boundary
condition). The rationale behind our finding is simple: the Chern number is by
definition an integral, whose integration domain shrinks to a single point in
the limit of a large supercell. 

\begin{figure}
\includegraphics[angle=-90,width=0.8\columnwidth]{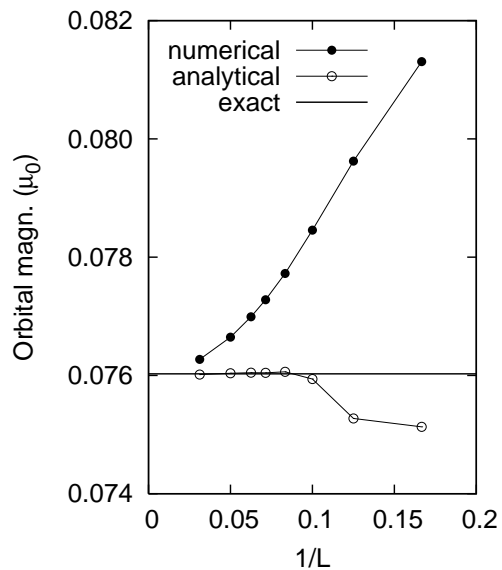}
\caption{Convergence of the orbital magnetization as a function of the supercell
size, evaluated using the single-point formulas (see text), for the Haldane
model Hamiltonian at $\phi = 0.4\pi$, The largest $L$ corresponds to 2048 sites.}
\label{fig:f3} \end{figure}

The single-point orbital magnetization $M$ of the model system, computed from
\eqs{mean}{mean4} as a function of the supercell size, is shown in Fig.~\ref{fig:f3}, again
for  $\phi = 0.4\pi$. In this case the analytical-derivative approach converges
definitely better, showing in fact the same kind of relative error as the Chern
number, while the numerical-derivative approach proves somewhat less accurate.

In conclusion, we provide here the key formulas for computing the orbital
magnetization of a condensed system from first principles in a supercell
framework and using a single ${\bf k}$ point, to be used as they stand within
Car-Parrinello simulations in an environment which breaks time-reversal
symmetry. We have validated the present formulas on a simple tight-binding model
Hamiltonian in 2d, and checked their (fast) convergence with the supercell
size. Last but not least, we have proved that even the Chern number---which
has a paramount relevance in quantum-Hall-effect simulations---can be computed
from a single Hamiltonian diagonalization, and converges fast with the supercell
size.

We acknowledge fruitful discussions with D. Vanderbilt and T. Thonhauser. Work
partly supported by  ONR through grant N00014-03-1-0570.

\section*{Appendix: More general boundary conditions}

The single-point formulas discussed so far are based on \equ{der}, with
$\lambda=1$, and eventually require diagonalizing the Hamiltonian at the
$\Gamma$ point only, ergo solving the Schr\"odinger equation with periodic
boundary conditions on the supercell. This is by far the most common choice
among Car-Parrinello practitioners, although other choices are possible.

In order to extend our single-point formulas to more general boundary
conditions it would be enough to switch from \equ{der} (at $\lambda = 1$) to
alternative expressions for the directional derivatives. The only
important requirement is that the two eigenstates therein differ by a
supercell reciprocal vector.

For the sake of simplicity we explicitly deal here only with the 2d case of
antiperiodic boundary conditions, corresponding to a zone-boundary single
point: in fact, antiperiodic eigenstates obtain by choosing the special
$\k$-vector $\kk_1 = ({\bf b}_1 + {\bf b}_2)/2$. It is then expedient to
define even $\kk_2 = ({\bf b}_1 - {\bf b}_2)/2 = \kk_1 - {\bf b}_2$ and to
switch from \equ{der} to \beq \ket{\partial_j u_{n0}} = \lim_{\lambda
\rightarrow 0} \; \frac{1}{2 \lambda |\kk_j|} ( \, \ket{u_{n \; \lambda
\kk_j}} - \ket{u_{n \; - \lambda \kk_j}} \, ) , \label{der1} \eeq where now
the $j$ subscript indicates the derivative in the direction of $\kk_j$. In
terms of such derivatives the magnetization formula, \equ{mean4} reads \beq M
= \frac{|\kk_1 \times \kk_2|}{c(2\pi)^2 |\kk_1| \, |\kk_2|} \,{\rm Im}\,
\bra{\partial_1 u_{n0}} (\, H_0 + \epsilon_{n0} -2 \mu \,)\,\ket{\partial_2
u_{n0}} , \label{mean5} \eeq and similarly for the Chern number.  

In the case of a large supercell  we approximate \equ{der1} with its
$\lambda=1$ value, noticing that all the needed states obtain from a single
Hamiltonian diagonalization at $\kk_1$. In fact $\ket{u_{n \; \kk_2}} =
\ei{{\bf b}_2 \cdot \r} \ket{u_{n \; \kk_1}}$ and $\ket{u_{n \; -\kk_j}} =
\ei{2 \kk_j \cdot \r} \ket{u_{n \; \kk_j}}$, where $2 \kk_j$ is a
reciprocal-lattice vector.  

While the states $\ket{u_{n \; \kk_1}}$ are to be used as they stand, the
states $\ket{u_{n \; -\kk_1}}$, $\ket{u_{n \; \kk_2}}$, and $\ket{u_{n \;
-\kk_2}}$ must be regularized to their dual counterpart, by means of the
obvious analogue of \equ{dual}. One then uses these states in \equ{der1}
with $\lambda=1$ and finally in \equ{mean5}.


\begin{thebibliography}{10}

\bibitem{KSV}
{ R. D. King-Smith and D. Vanderbilt, Phys. Rev. B {\bf 47}, 1651 (1993); D.
  Vanderbilt and R. D. King-Smith, Phys. Rev. B {\bf 48}, 4442 (1993)}.

\bibitem{rap_a12}
{ R. Resta, Rev. Mod. Phys. {\bf 66}, 899 (1994)}.

\bibitem{rap126}
{ R. Resta, D. Ceresoli, T. Thonhauser, and D. Vanderbilt, ChemPhysChem, {\bf
  6}, 1815 (2005)}.

\bibitem{Xiao05}
{ D. Xiao, J. Shi, and Q. Niu, Phys. Rev. Lett. {\bf 95}, 137204 (2005)}.

\bibitem{rap128}
{ T. Thonhauser, D. Ceresoli, D. Vanderbilt, and R. Resta, Phys. Rev. Lett.
  {\bf 95}, 137205 (2005)}.

\bibitem{rap130}
{ D. Ceresoli, T. Thonhauser, D. Vanderbilt, R. Resta, Phys. Rev. B {\bf 74},
  024408 (2006)}.

\bibitem{Xiao07}
{ D. Xiao, Y. Yao, Z. Fang, and Q. Niu, Phys. Rev. Lett. {\bf 97}, 026603
  (2007)}.

\bibitem{Shi07}
{ J. Shi, G. Vignale, D. Xiao, and Q. Niu,
  http:/$\!$/arxiv.org/abs/cond-mat/0704.3824}.

\bibitem{linear}
{ F. Mauri and S. G. Louie, Phys. Rev. Lett. {\bf 76}, 4246 (1996)}.

\bibitem{Mauri}
{ F. Mauri, B. G. Pfrommer, and S. G. Louie, Phys. Rev. Lett. {\bf 77}, 5300
  (1996); C. J. Pickard and F. Mauri, Phys. Rev. Lett. {\bf 88}, 086403
  (2002)}.

\bibitem{CP}
{ R. Car and M. Parrinello, Phys. Rev. Lett. {\bf 55}, 2471 (1985)}.

\bibitem{trois}
{ Sec. 8.3 in: R. Resta, {\it Berry Phase in Electronic Wavefunctions},
  Troisi\`eme Cycle Lecture Notes (Ecole Polytechnique F\'ed\'erale de
  Lausanne, Switzerland, 1996); also available at
  http://www-dft.ts.infn.it/\~{}resta/publ/notes\_trois.ps.gz}.

\bibitem{rap100}
{ R. Resta, Phys. Rev. Lett. {\bf 80}, 1800 (1998)}.

\bibitem{Thouless}
{ D. J. Thouless, {\it Topological Quantum Numbers in Nonrelativistic Physics}
  (World Scientific, Singapore, 1998)}.

\bibitem{DFT}
{ {\it Theory of the Inhomogeneous Electron Gas}, edited by S. Lundqvist and N.
  H. March (Plenum, New York, 1983)}.

\bibitem{Sai02}
{ N. Sai, K. M. Rabe, and D. Vanderbilt, Phys. Rev. B {\bf 66}, 104108 (2002)}.

\bibitem{Souza04}
{ I. Souza, J. \'I\~niguez and D. Vanderbilt, Phys. Rev. B {\bf 69}, 085106
  (2004)}.

\bibitem{Haldane88}
{ F. D. M. Haldane, Phys. Rev. Lett. {\bf 61}, 2015 (1988)}.

\bibitem{Yang96}
{ K. Yang and R. N. Bhatt, Phys. Rev. Lett. {\bf 76}, 1316 (1996)}.

\bibitem{Yang99}
{ K. Yang and R. N. Bhatt, Phys. Rev. B {\bf 59}, 8144 (1999)}.

\bibitem{Sheng03}
{ D. N. Sheng, X. Wan, E. H. Rezayi, K. Yang, R. N. Bhatt
  and F. D. M. Haldane, Phys. Rev. Lett. {\bf 90}, 256802 (2003)}.

\bibitem{Wan05}
{ X. Wan, D. N. Sheng, E. H. Rezayi, K. Yang, R. N. Bhatt,
  and F. D. M. Haldane, Phys. Rev. B {\bf 72}, 075325 (2005)}.

\end{thebibliography}
\end{document}